\documentstyle[preprint,eqsecnum,aps]{revtex}
\begin{document}
\draft
\preprint{slagter}
\title{Self-Gravitating Non-Abelian Cosmic String Solution}
\author{Reinoud J. Slagter \footnote{Electronic address: rslagt@sara.nl or
rjslagt@gironet.nl}}
\address{{\small\it Institute for Theoretical Physics, University of
Amsterdam, \\ Valckenierstraat 65, 1018XE Amsterdam, The Netherlands}}
\date{\today}
\maketitle
\begin{abstract}
{\it The coupled Einstein-Yang-Mills equations on a time dependent axially
symmetric spacetime are investigated, without imposing a priori any conditions
on the gauge field. There is numerical evidence for the existence of a regular
solution with the desired asymptotic features.
Just as in the supermassive abelian counterpart model, the formation of
a singularity at finite distance of the core of the string depends critically
on a parameter of the model, i.e., the constant value of one of the magnetic
components of the YM potentials.
The multiple-scale method could supply decisive answers concerning the
stability of the solution.}
\end{abstract}
%******************************************************************************
\section{Introduction}
After the astonishing numerical solution of Bartnik and McKinnon~\cite{BM}
of the static spherically symmetric Einstein-Yang-Mills equations with a
SU(2) gauge group, a fundamental question arose: do there exist essentially
non-abelian static globally regular solutions or blackholes, which carry
non-zero electric and/or magnetic charges~\cite{Biz,Zhou}?
In first instance, one could prove in the spherically symmetric case~\cite{Smol}
the existence of static solutions which are asymptotically flat and with finite
mass. The members of this family of solutions could be characterized by the
number of zeros of one of the  gauge potentials. In fact the repulsive Yang-Mills
force can balance the gravitational attractive force and prevent the formation
of singularities in spacetime. Later~\cite{Breit} the class of solutions were
extended to blackhole solutions with a horizon at some $r_h$ and oscillating
solutions which are not asymptotically flat. These non-abelian blackholes could
represent counter examples to the 'no-hair' conjecture.
Further, it was claimed~\cite{Lav} that both regular and blackhole solution in the
spherically symmetric case will have gravitational-like and
sphaleron-like instabilities.
When it was found that the BM solution
is unstable, a lot of researchers (see for example~\cite{Strau1,Brod}) tackled the general
stability problem of the coupled EYM system and
investigated the possible critical behavior of the solutions. It was realized
that they had much in common with the electroweak sphalerons.
In the spherical symmetric case the critical behavior was formulated
by Choptuik~\cite{Chop,Gun}. It occurred at the boundary in phase space
between initial data which eventually form a blackhole and data which do not.
Recently it was found, in a two dimensional Weinberg-Salam model with an axially
symmetric ansatz, that there is evidence for the existence of an electrically
charged sphaleron state~\cite{Saf}.
Here we consider the EYM system on an axially symmetric time dependent
spacetime and will compare our investigations
with the spherical symmetric case and the well known abelian cosmic string
solution~\cite{Gar,Lagu,Ortiz,Dy}, particularly the supermassive case.
In these U(1)-gauge cosmic string models it was found that as the energy scale
of symmetry breaking increases, the geometry around the string changes from
conical to an analog of a Kasner spacetime. But supermassive cosmic strings
may also arise at GUT scales if the coupling between scalar and gauge fields
is weak. Further, these low-energy supermassive strings are closely related to
global strings: they both show singular behavior at finite distance from the core
of the string.
We conjecture that the string-like solution in the Eintein-Yang-Mills model
shows some similar behavior.
The plan of this paper is as follows. In section II we derive the time dependent
equations for the Einstein-Yang-Mills system on an axially symmetric spacetime
using the algebraic manipulation
program MAPLE. In section III we consider the static situation in order to
gain insight into the asymptotic and singular behavior.
In section IV we solve the equations numerically and in section V we present a
conclusion and prospect concerning the stability. We conjecture that
conventional (linear) stability analysis
is inadequate to be applied to the situation where a singularity is formed.
The reason is that the oscillatory behavior of some of the gauge field components
are of high-frequency.
We suggest to apply the multiple-scale method~\cite{Taub,Cho,Sla3}, a method suitable
for handling high-frequency perturbations.
In a future work, we will extend the investigation to the spinning case, as
initiated before~\cite{Sla1,Sla2}.
%*******************************************************************************
\section{The field Equations}
Consider the Lagrangian of the SU(2) EYM system
\begin{equation}
S=\int d^4 x\sqrt{-det g}\Bigl[\frac{{\cal R}}{16\pi G}-\frac{1}{4}{\cal F}_{\mu\nu}^a
{\cal F}^{\mu\nu a}\Bigr],
\end{equation}
with the YM field strength
\begin{equation}
{\cal F}_{\mu\nu}^a =\partial_\mu A_\nu ^a-\partial_\nu A_\mu ^a +g\epsilon^{abc}
A_\mu^b A_\nu ^c,
\end{equation}
$g$ the gauge coupling constant, $G$ Newton's constant, $A_\mu^a$ the gauge
potential and ${\cal R}$ the curvature scalar.
The field equations then become
\begin{equation}
G_{\mu\nu}=-8\pi G{\cal T}_{\mu\nu},
\end{equation}
\begin{equation}
{\cal D}_\mu {\cal F}^{\mu\nu a}=0,
\end{equation}

with ${\cal T}$ the energy momentum tensor
\begin{equation}
{\cal T}_{\mu\nu}={\cal F}_{\mu\lambda}^a{\cal F}_\nu^{\lambda a}-\frac{1}{4}
g_{\mu\nu}{\cal F}_{\alpha\beta}^a {\cal F}^{\alpha\beta a}.
\end{equation}
We are interested in solutions on the cylindrical symmetric spacetime
\begin{equation}
ds^2=-e^{2(K-U)}(dt^2-dr^2)+e^{-2U}W^2d\varphi^2+e^{2U}dz^2,
\end{equation}
where $U, K$ and $W$ are functions of $t$ and $r$.
This cylindrical symmetric line element can formally be obtained from the
stationary axisymmetric line element used by Kleihaus and Kunz~\cite{Klei}
\begin{equation}
ds^2=-fdt^2+\frac{m}{f}(dr^2+dz^2)+\frac{lr^2}{f}d\varphi^2,
\end{equation}
with $f, m$ and $l$ functions of $r$ and $z$,
by the complex substitution $t\rightarrow iz$ and $z\rightarrow it $~\cite{Kram}.
Following Bais and Sasaki~\cite{Bais} we then can specify the gauge field potentials
$A_\mu = A_\mu^at_a$, where the $t_a$
are the anti-hermitian generators of the gauge group, by
\begin{equation}
A_\mu dx^\mu =A_0\hat\tau_\varphi dt +A_1\hat\tau_\varphi dr
+[\eta_2\hat\tau_r +\eta_1\hat\tau_z]dz+[\chi_2\hat\tau_r +\chi_1\hat\tau_z]d\varphi,
\end{equation}
with $A_i, \eta_i$ and $\chi_i$ functions of $t$ and $r$ and where
$\hat\tau_\varphi ,\hat\tau_r$ and $\hat \tau_z$ are the axial generators of
the SU(2) normalized such that $[\hat\tau_i,\hat \tau_j]=\epsilon_{ijk}\hat\tau_k$.
One can  reduce the number of gauge field potentials. On the Euclidean
spacetime, the self-duality condition eliminates two of the six functions~\cite
{Man}. One also can use the additional U(1) gauge freedom on the $A_\mu^a$,
$ A\rightarrow h^{-1}A h +h^{-1}dh$, with $h=exp[\psi(r,t)\hat\tau_\varphi]$
~\cite{Biz}. Here
we use the condition that on the spacetime (2.6), the energy-momentum component
${\cal T}_{z\varphi}$ must vanish. The most simple way to fulfil this condition
is $\eta_1=\chi_2$ and $\eta_2=\chi_1$. There are some other possibilities,
comparable with those found by Manton~\cite{Man}. However, these are quite
complicated as already noticed by Manton. So in our case, $T_{z\varphi}=0$
becomes
\begin{equation}
{\cal T}_{z\varphi}=g\eta_1(A_1^2-A_0^2)(1+2g\eta_2)+2(\partial_r\eta_1
\partial_r\eta_2-\partial_t\eta_1\partial_t\eta_2)+
A_0\partial_t\eta_2+A_1\partial_r\eta_2 =0.
\end{equation}
We then have two possibilities:
i. $\eta_2=$constant$=\eta_0$ and $A_0=\pm A_1$, or, ii. $\eta_2=\frac{-1}{2g}$.
We consider here case i.
The condition of case ii means that the YM mass scale $M_{YM}\equiv g\eta_0=
-\frac{1}{2}$. In section 3 we shall see that it corresponds in the static
case with a singular solution.

The differential equations for the several field variables become (from now
on we set $\eta_1 \equiv \eta$)
\begin{eqnarray}
\partial_t^2\eta-\partial_r^2\eta=&&\frac{1}{W}\partial_r\eta
(\partial_rW-2W\partial_rU)+\frac{1}{W}\partial_t\eta (2W\partial_tU-\partial_tW)
-g\eta_0(\partial_rA_0+\partial_tA_0)+ \cr &&\frac{g\eta_0A_0}{W}
(2W\partial_tU+2W\partial_rU-\partial_tW-\partial_rW)
+\frac{g\eta e^{2K}}{W^2}[\eta_0+g(\eta_0^2-\eta^2)],
\end{eqnarray}
\begin{eqnarray}
\partial_t^2A_0-\partial_r^2A_0=&&\frac{e^{2K}}{W^2}[g\eta_0(1-W^2e^{-4U})+1]
(\partial_t\eta -\partial_r \eta )\cr &&+(\partial_t A_0+\partial_r A_0)
(\frac{\partial_rW}{W}-\frac{\partial_tW}{W} +2\partial_rU-2\partial_tU
+2\partial_tK-2\partial_rK)\cr
&&-2\frac{e^{2K}A_0}{W^2}[g^2\eta^2+(1+g\eta_0)^2]
-2e^{2K-4U}A_0g^2(\eta^2+\eta_0^2),
\end{eqnarray}
\begin{equation}
\partial_r^2W-\partial_t^2W=-\frac{8\pi G}{W}\Bigl[e^{2K-2U}[g(\eta_0^2-\eta^2)
+\eta_0]^2+e^{-2K+2U}W^2(\partial_t A_0+\partial_rA_0)^2 \Bigr],
\end{equation}
\begin{eqnarray}
\partial_r^2K-\partial_t^2K=&&(\partial_tU)^2-(\partial_rU)^2+\frac{4\pi G}{W^2}
\Bigl[e^{-2U+2K}[g(\eta_0^2-\eta^2)+\eta_0]^2\cr
&&+W^2e^{2U-2K}(\partial_tA_0+\partial_rA_0)^2
-(e^{2U}-W^2e^{-2U})\Bigl((\partial_t\eta)^2-(\partial_r\eta)^2\Bigr)\cr
&&+2A_0[g\eta_0W^2e^{-2U}+(g\eta_0+1)e^{2U}](\partial_r\eta+\partial_t\eta)\Bigr]
\end{eqnarray}
and
\begin{eqnarray}
\partial_r^2U-\partial_t^2U=&&\partial_tU\partial_t\ln W-\partial_rU\partial_r \ln W
-\frac{4\pi G}{W^2}\Bigl[e^{-2U+2K}[g(\eta_0^2-\eta^2)+\eta_0]^2\cr
&&+W^2e^{2U-2K}(\partial_tA_0+\partial_rA_0)^2
+(e^{2U}-W^2e^{-2U})\Bigl((\partial_t\eta)^2
-(\partial_r\eta)^2\Bigr) \cr &&-2A_0[g\eta_0W^2e^{-2U}+(g\eta_0+1)e^{2U}]
(\partial_r\eta+\partial_t\eta)\Bigr].
\end{eqnarray}

From two combinations of the Yang-Mills equations (2.4), i.e.,
$$
[YM]_{\nu=\varphi,a=3} \pm \Bigl(\cos\varphi .[YM]_{\nu=z,a=1}+\sin\varphi .[YM]_{\nu=z,a=2}\Bigr),
$$
we obtain an expression for $A_0$,
\begin{equation}
A_0 =\frac{e^{2K-4U}W^2[\eta_0^2+g\eta_0(\eta_0^2-\eta^2)]
+e^{2K}[\frac{\eta_0}{g}(1+\eta_0 g)^2-\eta^2(1+g\eta_0)]}
{2\eta W[\partial_tW+\partial_rW-2W(\partial_rU+\partial_tU)]}
\end{equation}
and a first order equation for $A_0$
\begin{eqnarray}
\partial_tA_0+\partial_rA_0=-\frac{2A_0}{\eta}(\partial_t\eta+\partial_r\eta)
-\frac{e^{2K-4U}}{2\eta}[\eta_0^2+g\eta_0(\eta_0^2-\eta^2)] \cr
-\frac{e^{2K}}{2g\eta W^2}[g\eta^2(1+g\eta_0)-\eta_0(1+g\eta_0)^2].
\end{eqnarray}
In the abelian situation, the equations could be simplified, due to the
condition $T_{tt}=-T_{zz}$, and one obtains~\cite{Dy} $K=2U$. This is the
familiar self-gravitating Nielsen-Olesen vortex model studied first by
Garfinkle~\cite{Gar} in the static case.
In our case we obtain the condition
\begin{equation}
e^{-4U}W^2T_{zz}+T_{\varphi\varphi}-(T_{tt}-T_{rr})W^2e^{-2K}=0,
\end{equation}
which yields
\begin{equation}
\partial_r^2K-\partial_t^2 K=\partial_r^2U-\partial_t^2 U+
\frac{1}{W}(\partial_rU\partial_rW-\partial_tU\partial_tW)-
(\partial_rU)^2+(\partial_tU)^2-\frac{1}{W}(\partial_r^2W-\partial_t^2W).
\end{equation}
This relation can be used as constraint in the numerical code.
%******************************************************************************
%******************************************************************************
\section{The static case}
In order to obtain boundary conditions for the numerical integration of the
differential equations, we first consider the static situation.
The field equations reduce to
\begin{equation}
\partial_r^2W=\frac{-8\pi G}{W}\Bigl[e^{-2K+2U}W^2(\partial_rA_0)^2+
e^{2k-2U}[\eta_0+g(\eta_0^2-\eta^2)]^2\Bigr],
\end{equation}
\begin{eqnarray}
\partial_r[W\partial_r U]=&&-\frac{4\pi G}{W}\Bigl[
(W^2e^{-2U}-e^{2U})(\partial_r\eta)^2+W^2e^{2U-2K}(\partial_rA_0)^2
\cr &&-2[(g\eta_0+1)e^{2U}+g\eta_0W^2e^{-2U}]A_0\partial_r\eta
+e^{2K-2U}[\eta_0+g(\eta_0^2-\eta^2)]^2\Bigr],
\end{eqnarray}
\begin{equation}
\partial_r[W\partial_r K]=\frac{8\pi G}{W}\Bigl[
(g^2(\eta_0^2+\eta^2)A_0^2W^2e^{-2U}+e^{2U}[\partial_r\eta
+(1+g\eta_0)A_0]^2+g^2e^{2U}\eta^2A_0^2\Bigr],
\end{equation}
\begin{eqnarray}
\partial_r^2\eta =&&-\frac{1}{2}\partial_rA_0+
\frac{(1+2g\eta_0)A_0}{2W}(\partial_rW-2W\partial_rU)\cr
&&-\frac{g\eta}{2W^2}e^{2K}(W^2e^{-4U}+1)[\eta_0 +g(\eta_0^2-\eta^2)]
\end{eqnarray}
and
\begin{eqnarray}
\partial_r^2A_0=&&\partial_rA_0(2\partial_rK-\frac{\partial_rW}{W}-2\partial_rU)
+\frac{A_0}{W^2}e^{2K}[(1+g\eta_0)^2+g^2\eta^2]\cr
&&+g^2e^{2K-4U}A_0(\eta_0^2+\eta^2).
\end{eqnarray}
Further, (2.15) now reads
\begin{equation}
A_0 =\frac{e^{2K-4U}W^2[\eta_0^2+g\eta_0(\eta_0^2-\eta^2)]
+e^{2K}[\frac{\eta_0}{g}(1+g\eta_0 )^2-\eta^2(1+g\eta_0)]}
{2\eta W(\partial_rW-2W\partial_rU)}.
\end{equation}

From the condition that in  the static situation now $T_{tr}=0$, we obtain
\begin{equation}
\partial_r\eta=\frac{g^2A_0W^2(\eta_0^2+\eta^2)
+A_0e^{4U}[(1+g\eta_0)^2+g^2\eta^2]}{g\eta_0(W^2-e^{4U})-e^{4U}},
\end{equation}
which is consistent with one of the YM equations.
For the special case $g \eta_0=-\frac{1}{2}$ we obtain from (3.7)
\begin{equation}
\frac{\partial_r\eta}{(\frac{\eta}{\eta_0})^2+1}=-\frac{1}{2}A_0,
\end{equation}
which can be integrated together with (3.6) for some given spacetime.
However, we observe from (3.6) that $A_0$ becomes singular for $W\rightarrow
e^{2U}$. We then obtain from the YM equations just the condition
$g\eta_0=-\frac{1}{2}$.\\
The asymptotic features of the system can be analyzed by following
Garfinkle~\cite{Gar}. We write the stress tensor on an
orthonormal basis as
\begin{equation}
{\cal T}_{\mu\nu}=\sigma\hat t_\mu \hat t_\nu +{\cal P}_r \hat r_\mu \hat r_\nu
+{\cal P}_z \hat z_\mu\hat z_\nu +{\cal P}_\varphi \hat \varphi _\mu
\hat \varphi _\nu,
\end{equation}
where $\hat t^\mu = e^{-K+U}(\frac{\partial}{\partial t})^\mu, \hat r^\mu =
e^{-K+U}(\frac{\partial}{\partial r})^\mu ,\hat z^\mu =e^{-U}(\frac{
\partial}{\partial z})^\mu, \hat\varphi^\mu=\frac{e^U}{W}(\frac{\partial}
{\partial \varphi})^\mu$.
Using the conservation of stress energy
\begin{equation}
\nabla_\mu {\cal T}^{\mu\nu}=0,
\end{equation}
one obtains
\begin{equation}
\frac{d}{dr}\Bigl[W{\cal P}_r \Bigr]=(\sigma +{\cal P}_r +{\cal P}_z
-{\cal P}_\varphi )\Theta_1 -(\sigma+{\cal P}_r )\Theta_3 +{\cal P}_\varphi \Theta_2,
\end{equation}
where
\begin{eqnarray}
\Theta_1\equiv W\partial_r U,\quad \Theta_2\equiv \partial_r W,\quad
\Theta_3 \equiv W\partial_r K.
\end{eqnarray}
Let us assume that
\begin{equation}
\lim_{r\to \infty} 8\pi G W^2 e^{2K-2U}{\cal P}_r\rightarrow 0,
\end{equation}
and that $\Theta_i$ approach constant values as $r\rightarrow \infty$~\cite{Gar},
which are  fairly weak assumptions.
Then one can write with the help of (3.1)-(3.3)
\begin{equation}
\frac{d}{dr}\Bigl[8\pi G W^2 e^{2K-2U}{\cal P}_r\Bigr]=
\frac{d}{dr}(\Theta_2 \Theta_3-\Theta_1^2).
\end{equation}
One then obtains the asymptotic condition
\begin{equation}
\lim_{r\to\infty} (\Theta_2\Theta_3-\Theta_1^2)=0.
\end{equation}
So one has the two possibilities
\begin{eqnarray}
i:\quad &&W\partial_r U\vert_\infty =0,\quad W\partial_r W\partial_r K\vert_\infty =0\cr
ii:\quad &&(W\partial_r U\vert_\infty )^2=W\partial_r W\partial_r K)\vert_\infty.
\end{eqnarray}

If we denote by $W_\infty ,U_\infty$ and $K_\infty$ the values of the metric
fields far from the string, one obtains in the first case:
$U_\infty =a_1$, $W_\infty =a_2 r +a_3$ and $K_\infty =a_4$,
where the $a_i$ are constants.
The metric approaches in this case a conical spacetime~\cite{Lagu}
\begin{equation}
ds^2=e^{2(a_4 -a_1)}[-dt^2+dz^2]+e^{2a_1}dz^2+e^{-2a_1}(a_2r+a_3)^2d\varphi ^2].
\end{equation}

In the second case  one obtains in the special case $K=2U$ the Kasner-like
spacetime
\begin{equation}
ds^2=-(b_1 r+b_2)^4e^{2b_3}[dt^2-dr^2-dz^2]+\frac{e^{-2b_3}}{(b_1r+b_2)^2}d\varphi^2,
\end{equation}
with $b_i$ constants.
Now we can investigate the asymptotic behavior of $\eta$ and $A_0$ on the conical
spacetime (3.17) and compare the asymptotic features with the abelian counterpart
model.
Substituting the conical spacetime (3.17) into (3.6) and  (3.7) one obtains
solutions for $\eta$ and $A_0$. The solution is not sensitive for $a_1, a_3$ and $a_4$.
For positive $a_2$ the solution will possess singular behavior at
finite distance of the core, as expected from the abelian investigations of
Laguna and Garfinkle~\cite{Lagu}. In figure 1-3 we plotted $\eta$ and $A_0$
for $a_1=a_2=a_4=1$ and $a_3=2$ and for different values of $\eta_0$.
It is observed that the singularity is
pushed to infinity for smaller values of $\eta_0$ and $a_2>1$.
This kind of singular behavior is also encountered in the abelian Higgs
model~\cite{Lagu,Ortiz} when the scale of the symmetry breaking is far beyond the GUT scale, i.e.,
$8\pi G\nu^2 >> 10^{-6}$, where $\nu$ is the vacuum expectation value
of the Higgs field. In fact, the conical picture of the string   then fades away. It was
found that the metric of the transition between conical and
Kasner-like is cylindrical, i.e., $R^3\times S^1$. Further it was found
that supermassive strings can be formed at GUT scales if $\alpha\equiv
\frac{4\pi Ge^2}{\lambda}$, where e is the gauge coupling constant in the abelian
model and $\lambda$ the Higgs self coupling constant, is very small. The
singularity must then occur at a distance from the core of the string which
is many orders of magnitude greater than the present Hubble scale.
In our model we find a comparable dependency on $\eta_0$ of the behavior
of the singularity.
In general, however, one has to solve simultaneously the coupled system. Moreover,
one should make a distinction between the interior and exterior field equations
with proper matching conditions,
in order to get insight in the behavior close to the z-axis~\cite{Sla2}.

%******************************************************************************
%******************************************************************************
\section{ numerical solution}
The set of equation (2.10)-(2.14) can be solved numerically.
We used the ISML software package for numerically
solving coupled systems of nonlinear partial differential equations.
The package implements finite element
collocation methods based on piecewise polynomials for the spatial discretization
techniques. The time integration process is then accomplished for a set of
ordinary differential equations using banded Jacobians.
For the order of the piecewise polynomial space we took 5 and for the number of
subintervals into which the spatial domain is to be divided we took 11.
The relative error bound was $ 10^{-11}$.
In order to obtain from (2.15) a regular and asymptotically correct
initial value for $A_0$, we choose as initial values
$\eta(r,0)= e^{-r^2}+1$, $U(r,0)=K(r,0)=0$ and $W(r,0)=r+1$.
Figures 4, 5 and 6 show a typical regular solution of $A_0, \eta$ and
the metric component $e^{2K-2U}$ for some values of $\eta_0$ and $4\pi G$ and
suitable boundary conditions.
The behavior of $\eta$ remains regular everywhere.
In figure 7 we plotted $A_0$ for a smaller value of $\eta_0$.
We observe that  $A_0$ approaches asymptotically a constant value.
For runs where $A_0$ starts to oscillate, $e^{2K-2U}$ decreases
strongly, signalling the formation of a singularity.
However, a thoroughly investigation of the dependency on $g, \eta_0$
and $G$ will be necessary. This is currently under study.

%*****************************************************************************
%*****************************************************************************
\section{conclusions and outlook}
We investigated the existence of a possible solution of the coupled EYM system
on a time-dependent axially symmetric spacetime.
At least on the conical spacetime we find evidence for regular behavior of the
electric and magnetic components of the YM field for suitable values of the
parameters $\eta_0$, g and $4\pi G$. The formation of a singularity at finite
distance of the core of the string depends critically on $\eta_0$, the constant
value of one of the gauge field potentials, and $4\pi G$. Just as in the static
spherically symmetric EYMH case, where the ratio
$\frac{M_{YM}}{M_{Pl}}=\frac{g\nu}{\sqrt{4\pi G}}$($\nu$ is the vacuum expectation
value of the Higgs field)
plays a crucial role in the behavior of the solution, we find a crucial
dependency on the ratio $\frac{M_{YM}}{M_{Pl}}=\frac{g\eta_0}{\sqrt{4\pi G}}$.
In the abelian supermassive
model~\cite{Ortiz} a similar behavior is encountered where the singularity
arises at finite distance from the core of the string not only for large symmetry
breaking scales as found by~\cite{Lagu}, but also for  GUT scale. There is for fixed
symmetry breaking scale of order of the GUT scale a critical value for the
coupling of the scalar to the gauge field (i.e., $\frac{4\pi G e^2}{\lambda}$
with e the coupling constant and $\lambda$ the Higgs self coupling)
for which the singularity occurs at finite distance of the core.
These low energy solutions are more realistic because the larger the energy scale
the larger the angle deficit and  it no longer makes sense to talk about string
type solutions.
In our model, there will be a critical value $\frac{g\eta_0}{\sqrt{4\pi G}}$
for which the singularity is pushed to infinity.

In order to analyze the stability of the solutions, one usually linearizes
the field equations~\cite{Lav}, or one expands the field variables in a
physically unclear small parameter~\cite{Strau}.
One better can apply the so called multiple-scale (or two-timing) method,
developed decades ago by Taub~\cite{Taub} and Choquet-Bruhat~\cite{Cho}.
This method is is particularly
useful for constructing uniformly valid approximations to solutions
of perturbation problems. The idea is to expand the several field variables
in power series of the ratio of the characteristic wavelength of the perturbations
and the characteristic dimension of the background.
One  writes~\cite{Sla3}
\begin{eqnarray}
g_{\mu\nu}=&&\bar g_{\mu\nu}+\frac{1}{\omega}h_{\mu\nu}(x^\sigma;\xi)+
\frac{1}{\omega^2}k_{\mu\nu}(x^\sigma;\xi)+...\cr
A_\mu^a=&&\bar A_\mu^a +\frac{1}{\omega}B_\mu^a(x^\sigma;\xi)+\frac{1}{\omega^2}
C_\mu^a(x^\sigma;\xi)+...,
\end{eqnarray}
where $\xi\equiv \omega\Pi(x^\sigma )$ and $\Pi$ a phase function.
The parameter $\omega$ measures the ratio of the fast scale to the slow one.
The rapid variation only occur in the direction of the vector
$l_\sigma \equiv \frac{\partial \Pi}{\partial x^\sigma}$. For a function
$\Psi(x^\sigma ;\xi )$ one has
\begin{equation}
\frac{\partial\Psi}{\partial x^\sigma}= \partial_\sigma \Psi
+\omega l_\sigma \dot \Psi,
\end{equation}
where $\partial_\sigma\Psi \equiv \frac{\partial \Psi}{\partial x^\sigma}
\vert_{\xi fixed}$ and $\dot\Psi \equiv \frac{\partial \Psi}{\partial \xi}\vert
_{x^\sigma fixed}$.
Substituting the expansions of the field variables into the equations and
collecting terms of equal orders of $\omega$, one obtains propagation
equations for $\dot B_\mu^a$ and $\dot h_{\mu\nu}$ and 'back-reaction'
equations for $\bar h_{\mu\nu}$ and $\bar A_\mu^a$. It is clear from the
propagation equation that there will be a coupling between the high-frequency
gravitational field and the high-frequency behavior of $A_0$ when the
singularity will be approached. On his turn, $A_0$ will create a high-frequency
perturbation in $\eta$.

In a subsequent paper we will present this investigation.

%*****************************************************************************
%*****************************************************************************

\newpage
%******************************************************************************
%*******************************  FIGURES 1-3 *********************************
%******************************************************************************
\begin{figure}
\setlength{\unitlength}{1in}
\begin{picture}(4.5,4.5)
\put(0.,4.6){\special{em:graph rsfig1.pcx}}
\end{picture}
\caption{Plot of the gauge components $A_0$ and $\eta$ for $\eta_0 =0.5$.
The singularity appears when $\eta$ approaches zero}
\end{figure}
\newpage
\begin{figure}
\setlength{\unitlength}{1in}
\begin{picture}(4.5,4.5)
\put(0.,4.6){\special{em:graph rsfig2.pcx}}
\end{picture}
\caption{As figure 1, with $\eta_0 =0.2$. The singularity is encountered
at larger r value}
\end{figure}
\begin{figure}
\setlength{\unitlength}{1in}
\begin{picture}(4.5,4.5)
\put(0.,4.6){\special{em:graph rsfig3.pcx}}
\end{picture}
\caption{As figure 1, with  $\eta_0 =0.01$. The singularity is pushed to infinity}
\end{figure}

\newpage
%******************************************************************************
%**************************  FIGURES 4,5,6 and 7*******************************
\begin{figure}[h]
\setlength{\unitlength}{1in}
\begin{picture}(3.,3.)
\put(0.,3.0){\special{em:graph rsfig4.pcx}}
\end{picture}
\caption{{\it Plot of a long time-run of the gauge component $A0$ for g=-1,
$4\pi G=0.2$ and $\eta_0=0.2$}}
\end{figure}

\begin{figure}
\setlength{\unitlength}{1in}
\begin{picture}(3.,3.)
\put(0.,3.0){\special{em:graph rsfig5.pcx}}
\end{picture}
\caption{{\it Plot of $\eta$ for the situation of figure 4}}
\end{figure}

\begin{figure}
\setlength{\unitlength}{1in}
\begin{picture}(3.,3.)
\put(0.,3.0){\special{em:graph rsfig6.pcx}}
\end{picture}
\caption{{\it Plot of the metric component $e^{2K-2U}$
for the situation of figure 4}}
\end{figure}

\begin{figure}
\setlength{\unitlength}{1in}
\begin{picture}(3.,3.)
\put(0.,3.0){\special{em:graph rsfig7.pcx}}
\end{picture}
\caption{{\it As figure 4, but now for $\eta_0=0.01$}}
\end{figure}

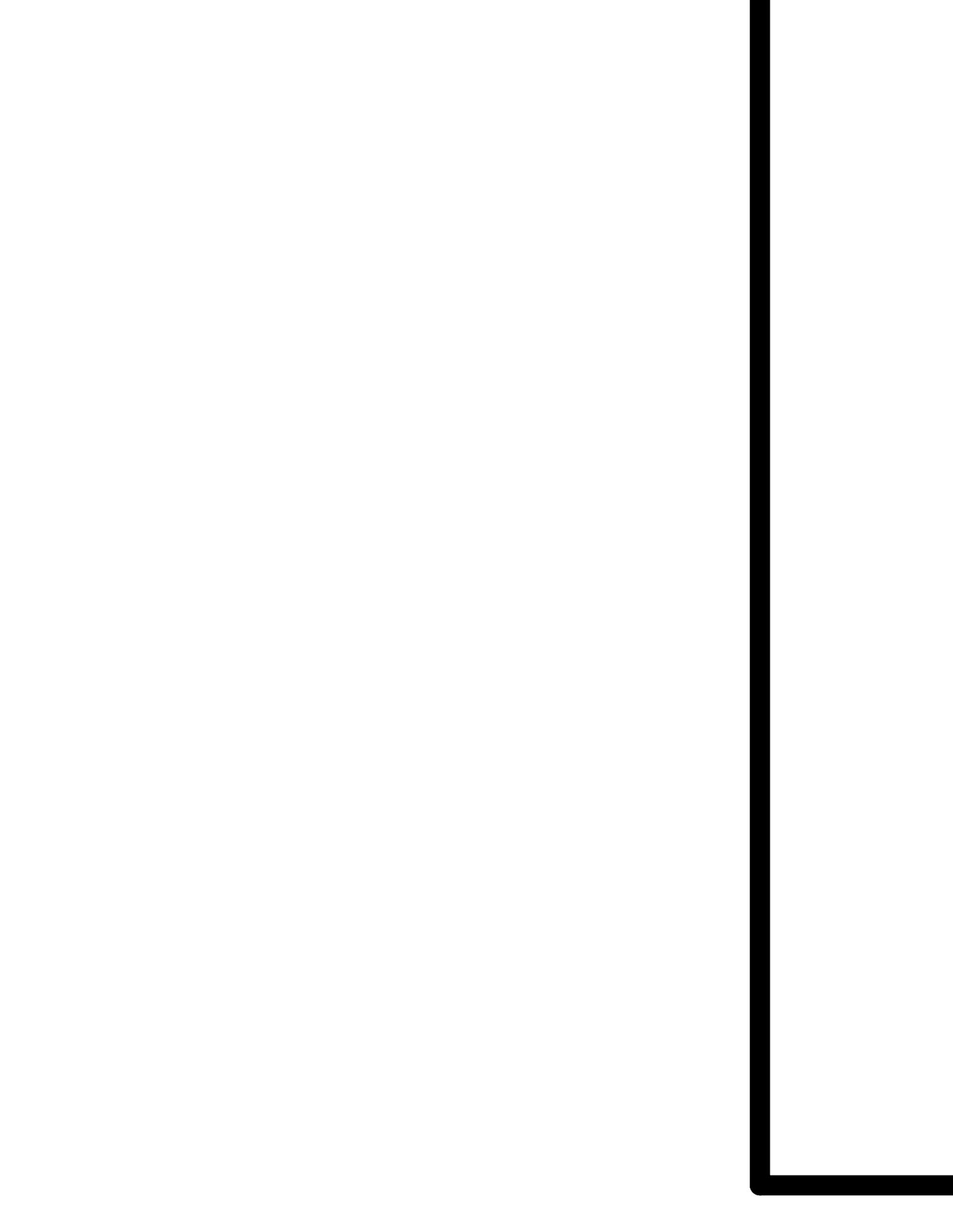
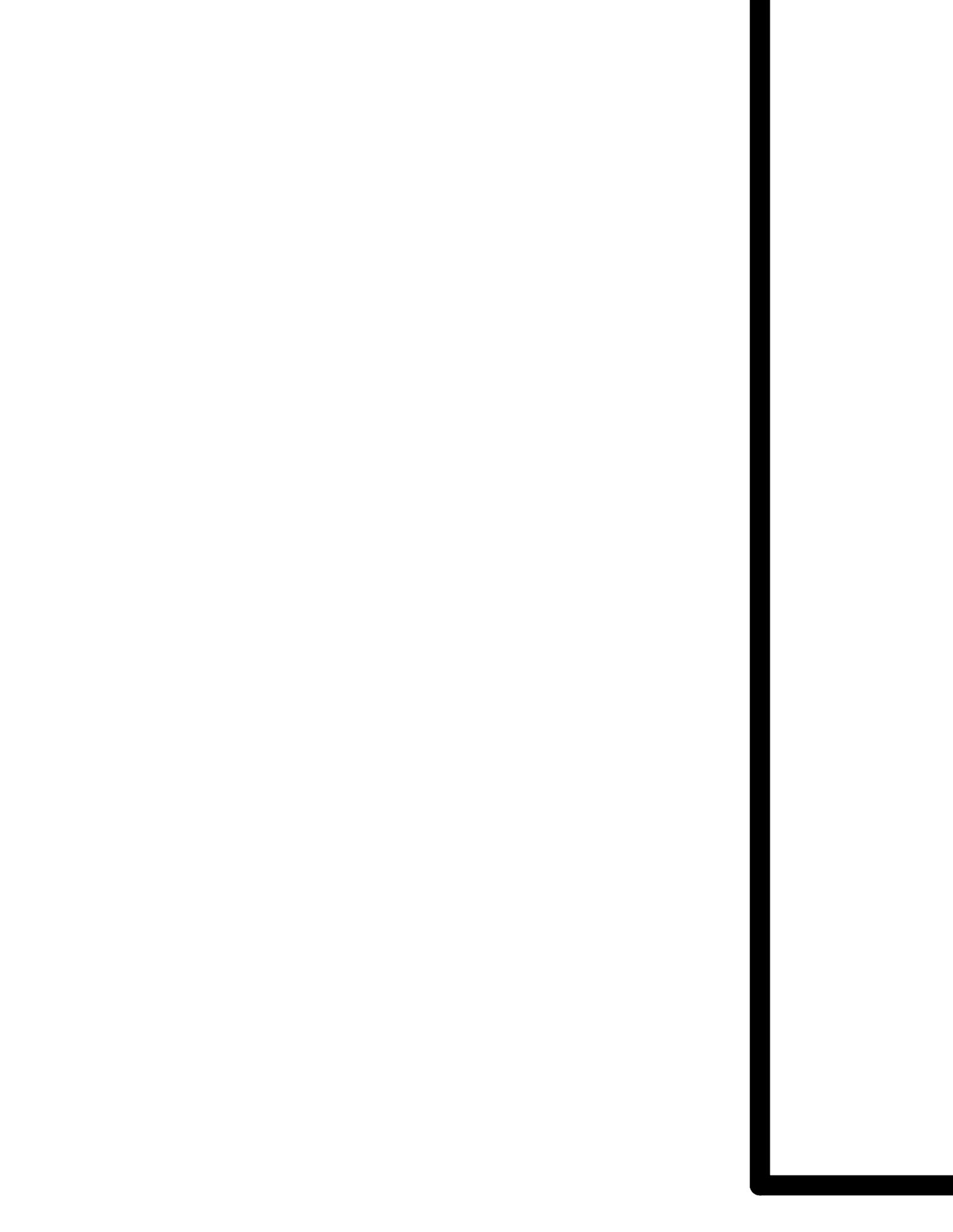
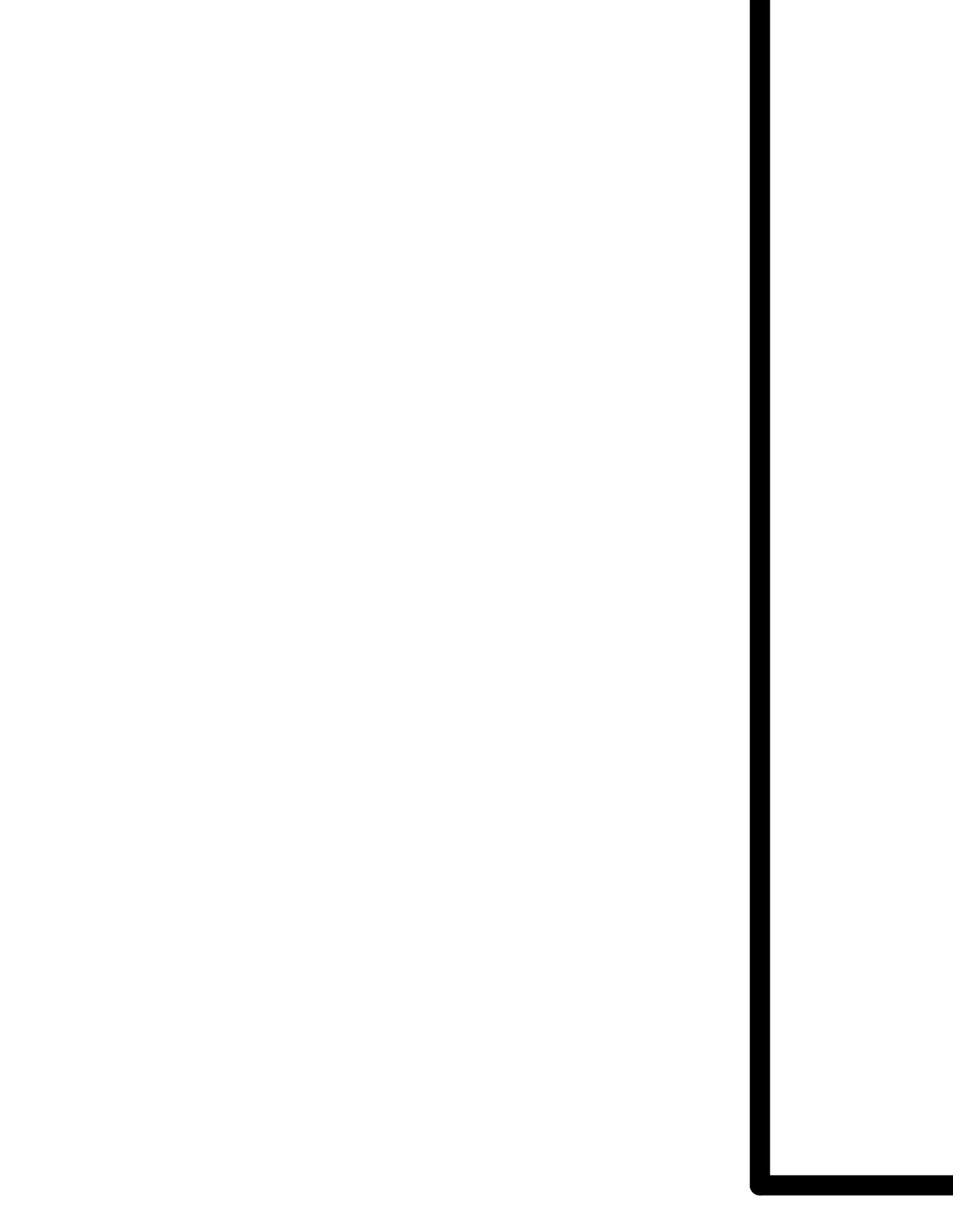
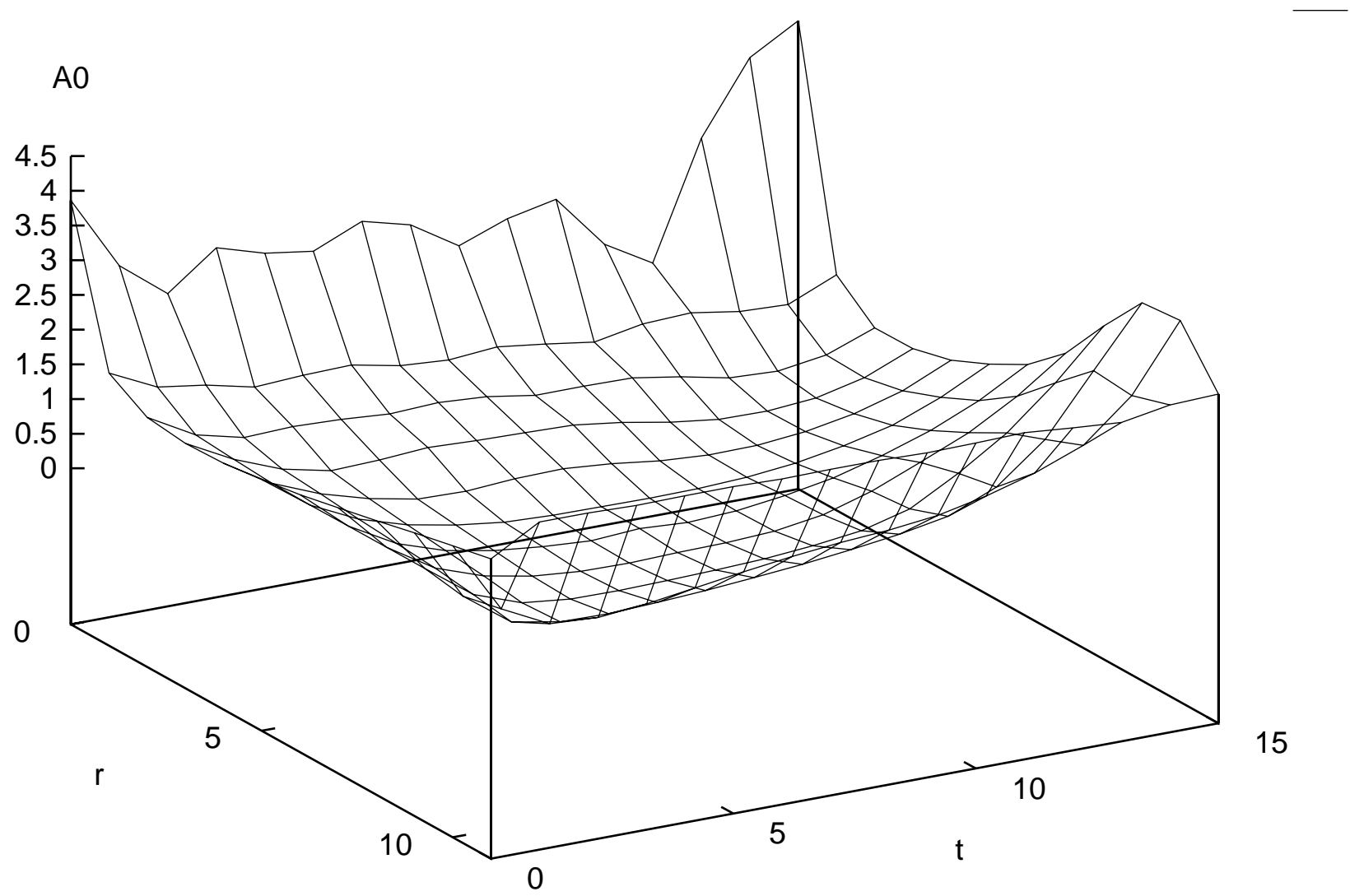
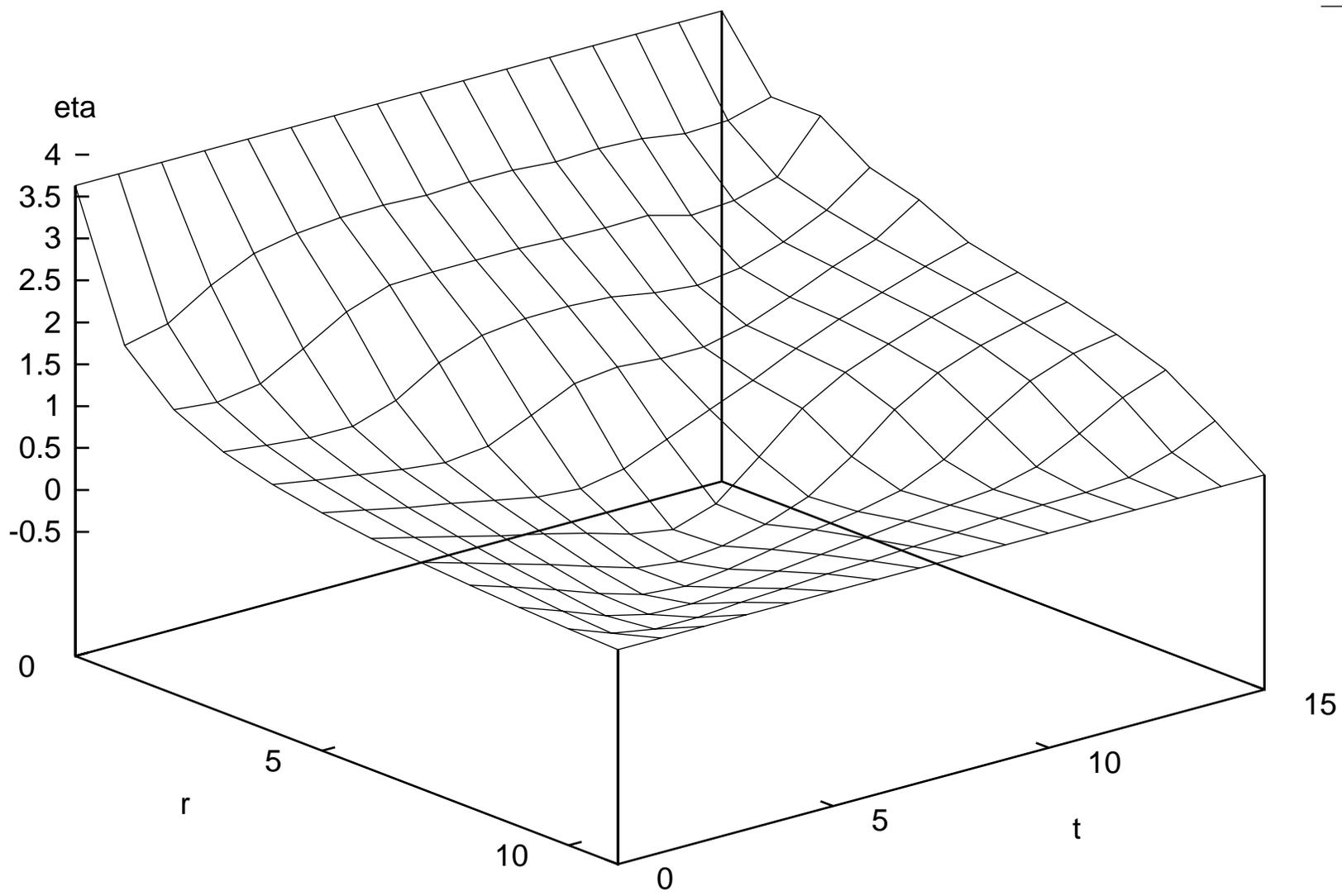
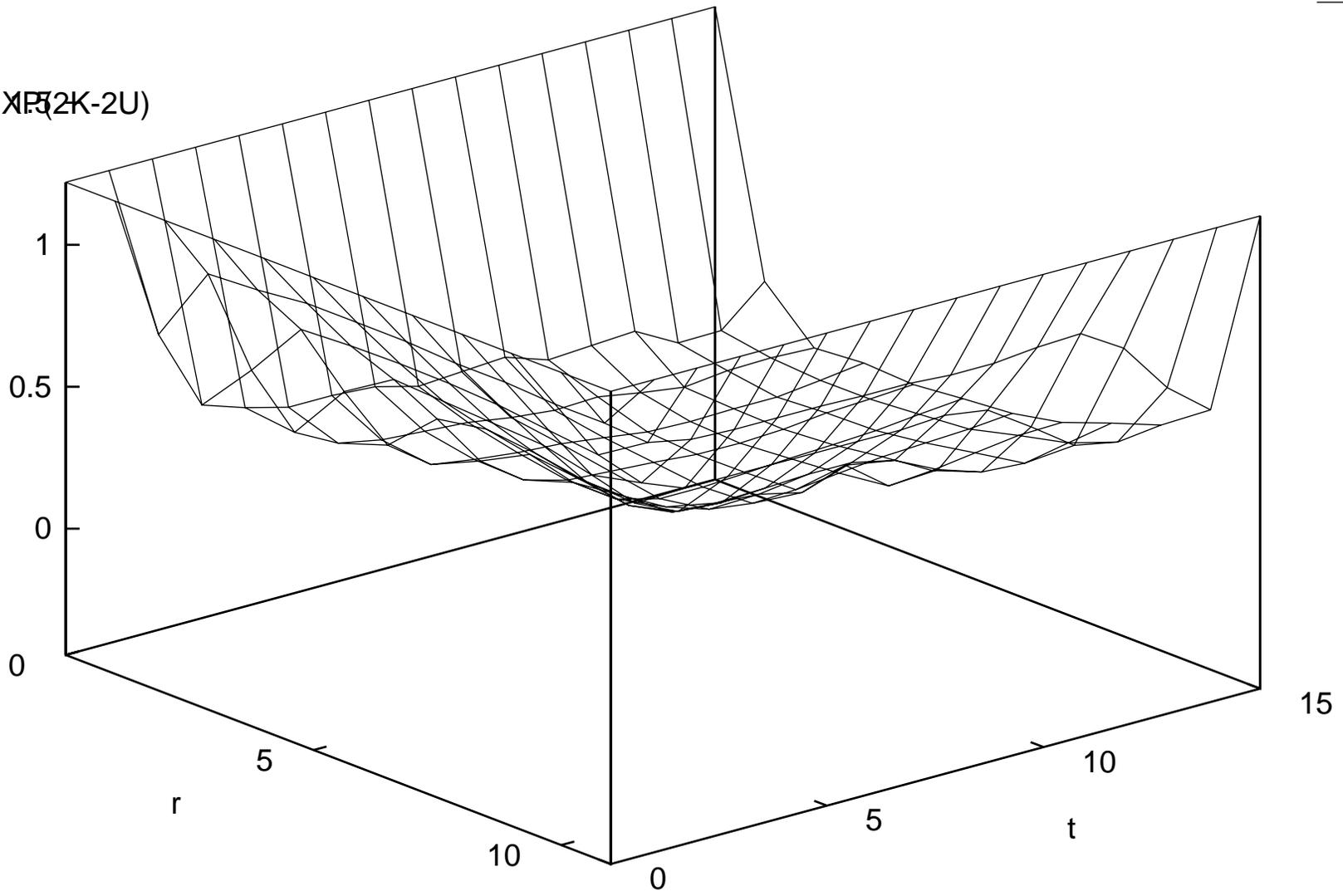
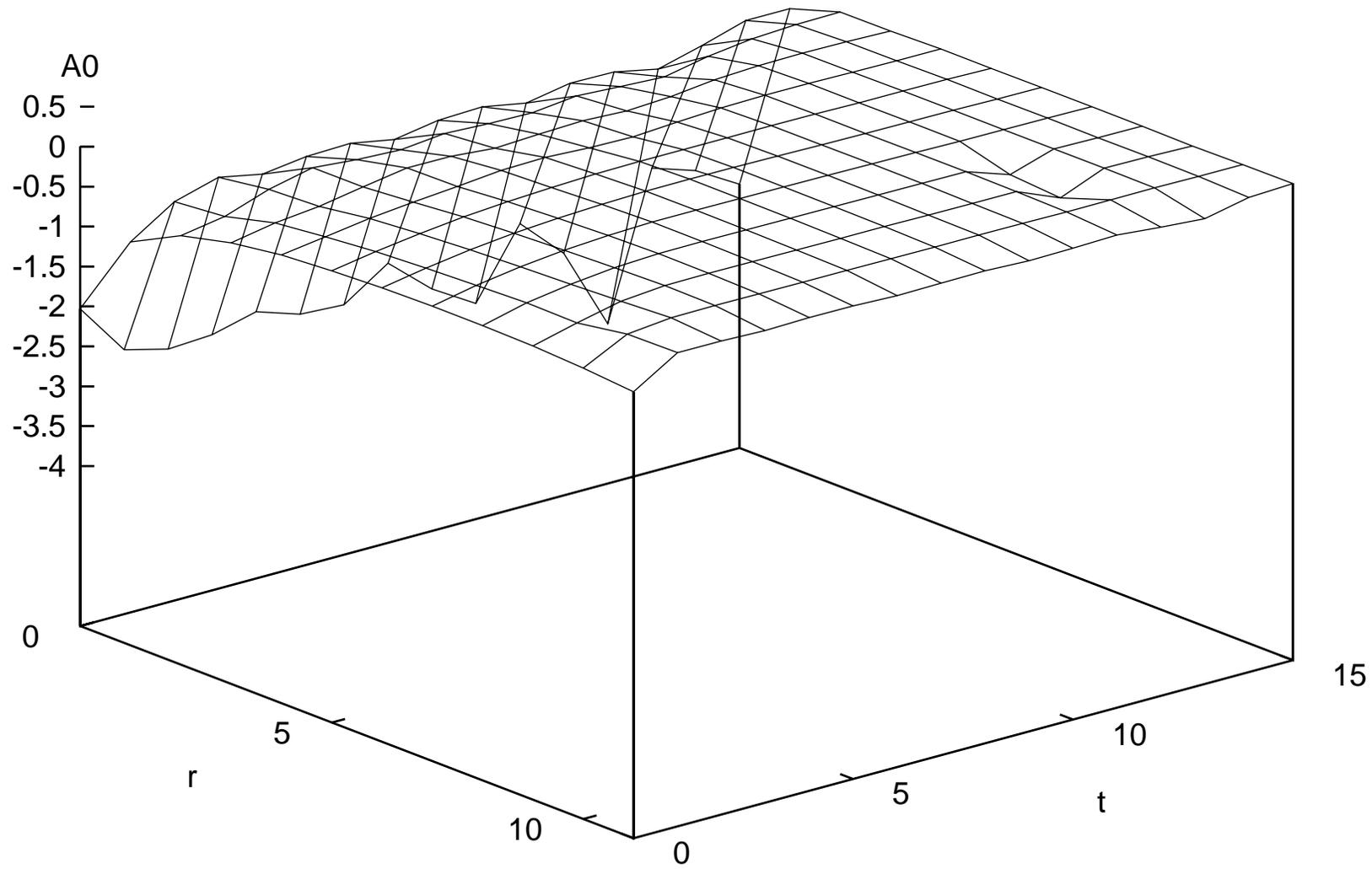
\end{document}